# The Principle of Least Action and Clustering in Cosmology

Mikel Susperregi[1] and James Binney[2]

[1] Astrophysics, 1 Keble Road, Oxford OX1 3RH, United Kingdom.
[2] Theoretical Physics, 1 Keble Road, Oxford OX1 3RH, United Kingdom.

**ABSTRACT**

A scheme is developed which enables one to trace backwards in time the cosmic density and velocity fields, and to determine accurately the current-epoch velocity field from the current-epoch density field, or vice versa. The scheme implements the idea of Giavalisco et al. (1993) that the principle of least action should be used to formulate gravitational instability as a two-point boundary-value problem. We argue that the Eulerian formulation of the problem is to be preferred to the Lagrangian one, on grounds of computational simplicity, of ease of interfacing with observational data, and of internal consistency at early times. The scheme is successfully tested on an exact solution in one dimension, and on currently Gaussian fields in one and two dimensions. The application of the scheme to real observational data appears to be eminently feasible, though computationally costly.

**Key words:** cosmology: theory – galaxies: clustering – large-scale structure of the universe.

## 1 INTRODUCTION

A key problem in cosmology is to understand and predict the evolution of the cosmic over-density and peculiar-velocity fields. Two propositions concerning these fields are widely accepted:

I. The evolution of the cosmic density and velocity fields on scales significantly larger than that of an individual galaxy has been dominated by gravity.

II. The cosmic density and velocity fields started from fluctuations of negligible amplitude at early times.

Proposition I enables us, in principle, to predict the currently observed fields that would have emerged from any given initial conditions. Conversely, it also enables us to determine the values of the fields at early times that are compatible with given fields at the current epoch. It follows that acceptance of both Propositions I and II is tantamount to placing strong constraints on the values of the fields at the current epoch. It turns out that these constraints leave only one of the fields free: once the density field has been chosen, the velocity field follows from the constraints, and vice versa. As was first pointed out by Peebles (1989), Proposition II in effect makes cosmology into a two-point boundary-value problem: observations provide us with a boundary condition at the current epoch, while Proposition II imposes a boundary condition at $t = 0$ from well-motivated physical assumptions.

Given that the current velocity and density fields are independently measurable, the capability to predict one field from the other provides a valuable empirical test of cosmogony. Moreover, if the two fields prove to be related in the way predicted by theory, it will be of interest to investigate the nature of the initial conditions. In particular, one would like to know whether the initial density field was Gaussian, as the simplest theory of the quantum origin of the cosmic density field in an inflationary universe predicts (e.g. Mukhanov et al. 1992; Peebles 1993).

Although these things are possible in principle, it is not at all clear how one should set about doing them in practice. One set of major complications arises from the fragmentary and error-prone nature of observational data. This paper is not concerned with such considerations, but concentrates instead on the dynamical problem of relating current-epoch data to vanishingly small initial density fluctuations.

Traditionally this problem has been approached through linear perturbation theory. Following Lifshitz (1946), one derives a linear o.d.e. for the evolution of the amplitude of each Fourier component $\delta_{\bm{k}}$ of the fractional over-density field $\delta(t,\bm{x})$. The general solution of this equation is a superposition of growing and decaying modes. Proposition II demands that the coefficient of the decaying mode be set equal to zero, with the result that the most general field compatible with Propositions I and II may at any time be represented as a sum of growing modes, with one arbitrary coefficient per mode. These coefficients precisely represent the independent degrees of freedom left to the fields by Propositions I and II. In fact, in linear theory, the velocity field is simply a well-determined multiple of the gradient of the Newtonian gravitational potential, so that the general velocity field may be expressed in terms of the arbitrary coefficients that determine the density field.

Unfortunately, at the current epoch the fields are significantly non-linear on scales smaller than $8h^{-1}$ Mpc and it is essential to have a theoretical framework that is valid in the non-linear regime. Zel'dovich (1970) made an important first step towards such a framework with his approximation

$$x(t,q) = x(0,q) - D(t)\nabla\Phi\big|_{\varpi(0,q)}, \qquad (1)$$

where $x(t)$ is the position at time $t$ of the particle $q$, $D(t)$ is the linear growth factor that emerges from Lifshitz' o.d.e., and $\Phi$ is the Newtonian potential. [It is conventional to label particles by their initial (Lagrangian) coordinates: one sets $q = x(0,q)$.] Numerical experiments (e.g. Efstathiou et al. 1985, Coles et al. 1993) have demonstrated that, in spite of its simplicity, the Zel'dovich approximation is remarkably accurate. Except for one-dimensional systems, it *is* only an approximation, however, and at late times it fails to mimic the formation of gravitational potential wells around collapsing structures because in it particles move in straight lines. Hence it underestimates the density contrast in regions where clustering is significant. Moreover, for a given velocity field it leads to two, mutually incompatible, estimates of the density field, one obtained from the continuity equation and the other from Euler's equation of conservation of momentum.

Several papers have recently proposed refinements to the Zel'dovich approximation that aim to extend the utility of the latter further into the non-linear regime (Nusser & Dekel 1992, Gramann 1993a,b, Buchert & Ehlers 1993). Perhaps because Zel'dovich obtained his approximation through inspired intuition rather than a systematic perturbation procedure, no unique refinement has emerged, although it has been shown that the Zel'dovich approximation belongs to a class of solutions in the lowest-order parameterization of perturbed particle orbits (Buchert 1992; Croudace et al. 1994).

Moutarde et al. (1991) enhanced the coordinate map (1) by writing

$$x(t,q) = q - D\nabla\Phi\big|_q + D^2 S^{(2)}(q), \qquad (2)$$

and this has been used to good effect by Bouchet et al. (1992) and Gramann (1993b), amongst others. Unfortunately, the algebra involved in these computations is laborious even at second order, and the prospect of calculating quantities to higher orders is not alluring. (For a detailed derivation of second and third order Lagrangian perturbation theory, see Buchert & Ehlers 1993.) Furthermore, these various perturbative approaches differ in their formalism and to a considerable extent in their predictions, with the result that appeal has been made to $n$-body simulations to resolve differences between them (e.g. Mancinelli et al. 1993).

More importantly, in these models, a solution of the governing equations is constructed as a Taylor series in the growth factor $D(t)$. The range of convergence of such a Taylor series is surely quite small because $\delta$ is known to become singular when the first caustics form; that is, for $D = D_0$, a thoroughly finite number, roughly equal to the inverse of the largest eigenvalue of the velocity deformation tensor. It would be surprising if the Taylor series for $\delta$ did not already break down when $D$ is quite a lot smaller than $D_0$. Hence, one is discouraged from pursuing the approaches of Bouchet et al. (1992) and Gramann (1993b) to higher orders not only by the considerable labour involved, but also by the fear that the higher-order terms, once obtained, would not prove meaningful at the current epoch.

The principle of least action is the natural device for solving a two-point boundary value problem such as that involved in seeking cosmic fields compatible with given observations and Propositions I and II. Not only does the least-action principle provide a natural mechanism for constraining the Universe to be smooth at early times, but it also ensures that expansions of the dynamical variables such as $\delta$ in powers of $D$ are constructed not as Taylor series but as least-squares approximations to the true solutions (see below). Since *any* function can be accurately approximated by a polynomial of sufficient degree, although very few functions have Taylor series with large radii of convergence, approximations to the dynamical variables obtained from the least-action principle are very much more likely to be useful than those obtained by a Taylor-series approach.

The principle of least action is most easily formulated for a system with a finite number of degrees of freedom, such as a collection of point masses or rigid bodies. Indeed, by treating each galaxy as a point mass, Peebles (1989, 1990) used the least-action principle to reconstruct the motion of the galaxies of the Local Group, while Dunn & Laflamme (1993) modelled neighbouring galaxies as extended, rigid objects to study the acquisition of angular momentum through tidal interactions, in reasonable agreement with the present observational estimates.

But at early times galaxies do not exist, and it cannot be appropriate to model the matter content of the Universe as a homogeneous distribution of massive bodies, whether pointlike or extended; until galaxies form, the only realistic formulation is in terms of a continuous field. If one adopts a Lagrangian approach, this field is the difference

$$\Psi(t,q) \equiv x(t,q) - q \qquad (3)$$

between the current and original positions of mass elements. In the Eulerian approach, the fields are the over-density and peculiar-velocity fields discussed above. Giavalisco et al. (1993) have formulated the problem for both types of field theory in the context of the principle of least action.

In principle, the Lagrangian approach is to be preferred because $\Psi$ remains finite at all times, whereas $\delta$ grows without limit at orbit-crossing time. Nevertheless, from the practical point of view the Eulerian approach enjoys several important advantages. First, its vector field $v$ is by Kelvin's theorem expressible as the gradient of a potential $\alpha$. Consequently, the Eulerian theory involves only two unrestricted scalar fields. Second, in the Eulerian theory the gravitational potential $\Phi$ is related to $\delta$ by Poisson's simple equation, whereas the relationship between $\Phi$ and $\Psi$ is complex in the extreme. Third, the current values of the Eulerian fields are more readily deduced from sparse and error-prone observations than are the Lagrangian displacement vectors $\Psi_i$ of galaxies, with the result that the Eulerian theory is easier to apply to observational data than the Lagrangian theory would be. Finally, in Eulerian theory one may readily study the tidal forces on any given comoving region by calculating the shear and tidal tensors, whereas this undertaking is difficult in the Lagrangian framework.

For these reasons, we have chosen to implement the Eulerian scheme that was outlined by Giavalisco et al. (1993). Section 2 sets out the basic formalism. Section 3 applies this formalism to an exactly soluble one-dimensional model and examines the results for some particular realizations of Gaussian random fields. Section 4 presents results for two-dimensional fields. Section 5 explains the relation of the present scheme to similar schemes for following the evolution of the cosmic fields. Section 6 sums up and describes how the scheme might be applied to real data.

## 2 EULERIAN RECONSTRUCTION

Giavalisco et al. (1993) show that the standard equations governing the evolution of cosmological perturbations in $d$ dimensions may be derived from the action

$$S = \int dt \int d^d x \, \mathcal{L}, \tag{4}$$

where $x$ are comoving coordinates and the Lagrangian density $\mathcal{L}$ is given by

$$\mathcal{L} = \tfrac{1}{2}(1+\delta)\,v^2 - \Phi\delta + \alpha\left\{\dot{\delta} + \tfrac{1}{a}\nabla\cdot[(1+\delta)v]\right\} - \frac{|\nabla\Phi|^2}{8\pi G\rho_b a^2}. \tag{5}$$

Here $\delta$ and $\Phi$ are the fractional over-density and associated Newtonian potential, $v$ is the peculiar velocity field, $a(t)$ is the expansion factor of the universe and $\rho_b(t)$ is the unperturbed density. Varying the action with respect to $v$ and the potential $\Phi$ we find

$$\begin{aligned} v(t,x) &= \tfrac{1}{a}\nabla\alpha(t,x), \\ \nabla^2\Phi(t,x) &= 4\pi G\rho_b a^2 \delta(t,x). \end{aligned} \tag{6}$$

The first of these equations tells us that $\alpha$ is the velocity potential, while the second is simply Poisson's equation. Varying the action (4) with respect to $\alpha$ and $\delta$, and then eliminating $v$ with the aid of the first of equations (6) yields

$$\begin{aligned} 0 &= \dot{\delta} + \tfrac{1}{a^2}\nabla\cdot[(1+\delta)\nabla\alpha], \\ 0 &= \dot{\alpha} + \tfrac{|\nabla\alpha|^2}{2a^2} + \Phi. \end{aligned} \tag{7}$$

These are the continuity and Bernouilli equations. By discarding the quadratic terms in equations (7), we recover the equations of linear theory.

When $v$ is similarly eliminated between equations (5) and (6), $\mathcal{L}$ becomes

$$\mathcal{L} = \alpha\dot{\delta} - \frac{(1+\delta)|\nabla\alpha|^2}{2a^2} - \Phi\delta - \frac{|\nabla\Phi|^2}{8\pi G\rho_b a^2}. \tag{8}$$

This form demonstrates that $\alpha$ and $\delta$ are mutually conjugate variables.

The solutions to the field equations (7), are constrained by the mixed boundary conditions

$$\delta(0,x) = 0 \quad \text{and} \quad \delta(t_0,x) = \delta_0(x), \tag{9}$$

where $t_0$ is the present time and $\delta_0(x)$ is a function that may, in principle, be determined observationally. We express the fields as Fourier integrals of the form

$$\begin{aligned} \delta(t,x) &= \sum_{n=0}^{\infty} f_n(t) \int \frac{d^d k}{(2\pi)^d}\, \delta_{k,n}\, e^{i k\cdot x}, \\ \Phi(t,x) &= 4\pi G\rho_b a^2 \sum_{n=0}^{\infty} f_n(t) \int \frac{d^d k}{(2\pi)^d}\, \Phi_{k,n}\, e^{i k\cdot x}, \\ v(t,x) &= a \sum_{n=0}^{\infty} g_n(t) \int \frac{d^d k}{(2\pi)^d}\, v_{k,n}\, e^{i k\cdot x}, \\ \alpha(t,x) &= a^2 \sum_{n=0}^{\infty} g_n(t) \int \frac{d^d k}{(2\pi)^d}\, \alpha_{k,n}\, e^{i k\cdot x}, \end{aligned} \tag{10}$$

where

$$\begin{aligned} f_n(t) &\equiv D(t)\bigl[D(t)-1\bigr]^n, \\ g_n(t) &\equiv \dot{D}(t)\bigl[D(t)-1\bigr]^n, \end{aligned} \quad n=0,1,2,\ldots \tag{11}$$

with $D(t_0) = 1$. In view of (6), we have

$$\begin{aligned} v_{k,n} &= i k\,\alpha_{k,n}, \\ \Phi_{k,n} &= -\delta_{k,n}/k^2, \end{aligned} \tag{12}$$

so the coefficients $\delta_{k,n}$ and $\alpha_{k,n}$ determine all four fields. Notice that $\delta_{k,0}$ and $\alpha_{k,0}$ are the Fourier transforms of the current-epoch fields rather than of the early-time fields as in conventional perturbative cosmology (e.g. Padmanabhan 1993; Jain & Bertschinger 1994). Notice also that the $\delta_{k,n}$ and $\alpha_{k,n}$ do not depend on time, the time-dependence of the physical fields having been absorbed into the functions $f_n(t)$ and $g_n(t)$ defined by equations (11).

Substituting the expansions (10) into equations (7), multiplying the equations through by $g_r$ and $f_r$, respectively, and integrating over time, we obtain a nonlinear system of time-independent equations for the $\delta_{k,n}$ and $\alpha_{k,n}$:

$$\begin{aligned} \sum_{n=0}^{\infty} \langle \dot{f}_n g_r\rangle \delta_{k,n} - k^2 \sum_{n=0}^{\infty} \langle g_n g_r\rangle \alpha_{k,n} \\ = \sum_{n,m=0}^{\infty} \langle f_n g_m g_r\rangle \int \frac{d^d p}{(2\pi)^d}\, k\cdot p\, \delta_{k-p,n}\,\alpha_{p,m} \end{aligned} \tag{13}$$

and,

$$\begin{aligned} 4\pi G \sum_{n=0}^{\infty} \langle \rho_b f_n f_r\rangle \delta_{k,n} - k^2 \sum_{n=0}^{\infty} \langle (\dot{g}_n + 2\tfrac{\dot{a}}{a} g_n) f_r\rangle \alpha_{k,n} \\ = -\tfrac{1}{2} k^2 \sum_{n,m=0}^{\infty} \langle g_n g_m f_r\rangle \times \\ \times \int \frac{d^d p}{(2\pi)^d}\, p\cdot(k-p)\,\alpha_{k-p,n}\,\alpha_{p,m}, \end{aligned} \tag{14}$$

where

$$\langle x\rangle \equiv \int_0^{t_0} dt\, a^2(t) x(t). \tag{15}$$

The values of the angle brackets required in (13), and (14) are calculated in Appendix A for the case $\Omega = 1$, $\Lambda = 0$. The corresponding values for other cosmological models are easily calculated. For concreteness, we shall henceforth assume that $\Omega = 1$, $\Lambda = 0$.

Equations (13) and (14) provide a pair of equations for each wavevector $k$ and non-negative integer $r$. Thus, for every $k$ there is a infinite system of equations $r = 0, 1, 2, \ldots$. We simplify this system by (i) limiting the index $r$ to $r \leq N$, and (ii) assuming that the fields satisfy periodic boundary conditions in a box of side $L$. This assumption enables us to replace the Fourier integrals (10) by sums over a countable set of wavenumbers $k_i$. These sums we truncate after $K$ wavevectors in each coordinate direction. With these restrictions and approximations we have

$$\begin{aligned} \delta(t,x) &= \frac{1}{L^d} \sum_{k} e^{i k\cdot x} \sum_{n=0}^{N} f_n(t)\,\delta_{k,n}, \\ \alpha(t,x) &= \frac{a^2(t)}{L^d} \sum_{k} e^{i k\cdot x} \sum_{n=0}^{N} g_n(t)\,\alpha_{k,n}, \end{aligned} \tag{16}$$

where the $\delta_{\boldsymbol{k},n}$ and $\alpha_{\boldsymbol{k},n}$ satisfy (see Appendix B for intermediate results),

$$\sum_{n=0}^{N} \phi(n+r)\left(\frac{2r-3n}{2(n+r)}\delta_{\boldsymbol{k},n} - k^2\alpha_{\boldsymbol{k},n}\right)$$
$$= -\frac{5}{2L^d}\sum_{n,m=0}^{N}\eta(n+m+r)\sum_{\boldsymbol{p}} \boldsymbol{k}\cdot\boldsymbol{p}\,\delta_{\boldsymbol{k}-\boldsymbol{p},n}\,\alpha_{\boldsymbol{p},m}, \quad (17)$$

and

$$\sum_{n=0}^{N} \phi(n+r)\left(\delta_{\boldsymbol{k},n} - \frac{3r-2n}{3(n+r)}k^2\alpha_{\boldsymbol{k},n}\right) = \frac{5k^2}{6L^d}\times$$
$$\times \sum_{n,m=0}^{N}\eta(n+m+r)\sum_{\boldsymbol{p}} \boldsymbol{p}\cdot(\boldsymbol{k}-\boldsymbol{p})\,\alpha_{\boldsymbol{k}-\boldsymbol{p},n}\,\alpha_{\boldsymbol{p},m}. \quad (18)$$

Here

$$\phi(n) \equiv (-4)^n\frac{n!\,(n+2)!}{(2n+5)!},$$
$$\eta(n) \equiv \frac{\phi(n+1)}{n+1}. \quad (19)$$

For concreteness we restrict ourselves to the case in which the current-epoch density field is specified – the case of specified velocity potential is entirely analogous and requires only rearrangement of the coefficients. In this case it is expedient to arrange equations (17) and (18) such that the non-linear terms and terms involving $\delta_{\boldsymbol{k},0}$ appear on the right. Then for each wavevector $\boldsymbol{k}$ we have the $(2N+1)$-dimensional matrix equation

$$\begin{pmatrix}\boldsymbol{D}^{(1)} & \boldsymbol{A}\\ \boldsymbol{A}^T & \boldsymbol{D}^{(2)}\end{pmatrix}\begin{pmatrix}\boldsymbol{\Omega}(k)\\ \boldsymbol{\Delta}(k)\end{pmatrix} = \begin{pmatrix}\boldsymbol{C}^{(1)}(k)\\ \boldsymbol{C}^{(2)}(k)\end{pmatrix} - \delta_{\boldsymbol{k},0}\begin{pmatrix}\boldsymbol{\Psi}\\ \boldsymbol{\Psi}\end{pmatrix}, \quad (20)$$

where the vectors to be calculated are,

$$\boldsymbol{\Omega}(k) \equiv -k^2\begin{pmatrix}\alpha_{\boldsymbol{k},0}\\ \alpha_{\boldsymbol{k},1}\\ \alpha_{\boldsymbol{k},2}\\ \vdots\\ \alpha_{\boldsymbol{k},N}\end{pmatrix}, \quad \boldsymbol{\Delta}(k) \equiv \begin{pmatrix}\delta_{\boldsymbol{k},1}\\ \delta_{\boldsymbol{k},2}\\ \delta_{\boldsymbol{k},3}\\ \vdots\\ \delta_{\boldsymbol{k},N}\end{pmatrix}, \quad (21)$$

and the elements of the submatrices are,

$$A_{rn} \equiv \phi(r+n),$$
$$D_{rn}^{(1)} \equiv \frac{3r-2n}{3(n+r)}\phi(r+n), \quad (22)$$
$$D_{rn}^{(2)} \equiv \frac{2r-3n}{2(n+r)}\phi(r+n).$$

The quadratic subcolumns on the right-hand side of (20) are given by,

$$C_r^{(1)} \equiv \frac{5k^2}{6L^d}\sum_{\substack{n,m=0\\ \boldsymbol{p}}}^{N}\eta(n+m+r)\,\boldsymbol{p}\cdot(\boldsymbol{k}-\boldsymbol{p})\,\alpha_{\boldsymbol{k}-\boldsymbol{p},n}\,\alpha_{\boldsymbol{p},m},$$
$$C_r^{(2)} \equiv -\frac{5}{2L^d}\sum_{\substack{n,m=0\\ \boldsymbol{p}}}^{N}\eta(n+m+r)\,\boldsymbol{k}\cdot\boldsymbol{p}\,\delta_{\boldsymbol{k}-\boldsymbol{p},n}\,\alpha_{\boldsymbol{p},m}, \quad (23)$$

while

$$\Psi_r \equiv \phi(r) \quad (24)$$

independent of $\boldsymbol{k}$.

The $\boldsymbol{C}^{(i)}$ couple the matrix equations of the set (20) for different wavevectors $\boldsymbol{k}$. So one would ideally solve (20) as a set of $K^d$ equations in $K^d$ $(2N+1)$-dimensional variables, using the Newton-Raphson algorithm to handle the non-linearity inherent in the $\boldsymbol{C}^{(i)}$. For a realistic number $K$ of wavevectors, this approach is unfeasible, however, and we have solved the system as follows.

Given the Fourier coefficients $\delta_{\boldsymbol{k},0}$ of a current-epoch density field, we estimate $\alpha_{\boldsymbol{k},0}$ from standard linear theory:

$$\frac{\dot D}{D}\delta \simeq \dot\delta \simeq -\frac{\nabla^2\alpha}{a^2} \Rightarrow \delta_{\boldsymbol{k},0} \approx k^2\alpha_{\boldsymbol{k},0}. \quad (25)$$

Then we set $\delta_{\boldsymbol{k},n} = \alpha_{\boldsymbol{k},n} = 0$ for $0 < n \leq N$ and evaluate the non-linear terms $\boldsymbol{C}^{(i)}$. The r.h.s. of (20) may now be completely evaluated and $\boldsymbol{\Omega}(k), \boldsymbol{\Delta}(k)$ solved for by multiplying (20) through by the inverse of the matrix on its l.h.s. The values of $\delta_{\boldsymbol{k},n}$ and $\alpha_{\boldsymbol{k},n}$ for $n \neq 0$ recovered in this way are used to re-evaluate the $\boldsymbol{C}^{(i)}$ and the procedure iterated. One finds that the iterated values of the fields converge to a solution of the equations provided the initially specified fields are not large, and therefore the non-linear terms are of only secondary importance.

If the initial fields are large, brute-force iteration of the type just described does not yield a solution, and we proceed as follows. We use brute-force iteration to solve the equations obtained by replacing the $\boldsymbol{C}^{(i)}$ by $m^{-1}\boldsymbol{C}^{(i)}$, where $m$ is a sufficiently large integer. Then we use the solution we have just obtained as the starting point for an iterative solution of the equations obtained by replacing $\boldsymbol{C}^{(i)}$ by $2m^{-1}\boldsymbol{C}^{(i)}$. For $m$ sufficiently large, these iterations also converge, enabling the procedure to be repeated with $3m^{-1}\boldsymbol{C}^{(i)}$ on the r.h.s., and so on until equations (20) have been solved.

## 3 APPLICATION TO ONE-DIMENSIONAL SYSTEMS

### 3.1 An analytically soluble model

Equations (20) simplify greatly in the case $d = 1$. Moreover, in $d = 1$ we have an exact solution to the perturbative cosmological problem against which to test our apparatus. Namely the plane-wave solution (e.g. Padmanabhan 1993)

$$\delta(t,x) = \frac{(t/\tau)^{2/3}\cos(\kappa q)}{1 - (t/\tau)^{2/3}\cos(\kappa q)}, \quad (26)$$

where the Eulerian coordinate $x$ is related to the comoving Lagrangian coordinate $q$ by

$$x(t,q) = q - \left(\frac{t}{\tau}\right)^{2/3}\frac{\sin(\kappa q)}{\kappa}. \quad (27)$$

As $t \to 0$, the density field (26) tends to a sinusoidal wave. We investigate the ability of our machinery to recover this from the non-sinusoidal field described by (26) for $t \simeq \tau$.

At $t_0$ the density contrast $\delta_0$ given by (26) satisfies

$$-\frac{(t_0/\tau)^{2/3}}{1 + (t_0/\tau)^{2/3}} \leq \delta \leq \frac{(t_0/\tau)^{2/3}}{1 - (t_0/\tau)^{2/3}}. \quad (28)$$

Hence, for $t_0 = \tau/2^{3/2} \simeq 0.35\tau$, we have that $\delta_0$ lies between $\delta_{\max} = 1$ and $\delta_{\min} = -1/3$. Evidently, nonlinear effects are important from at least this time onwards.

The full curves in Fig. 1 show the analytical density field at $t = 0$ (thin curve) and $t = 0.46\tau$ (thick curve), while the dashed curve shows the initial density field obtained by applying the least-action principle to the field at time $t_0 = \frac{1}{4}\tau$. Fig. 2 shows the corresponding velocity fields. Note that the quantities plotted are $\delta/D$ and $v/a\dot{D}$, so any evolution apparent in these figures is due to non-linearity; the growth in amplitudes predicted by linear theory has been factored out.

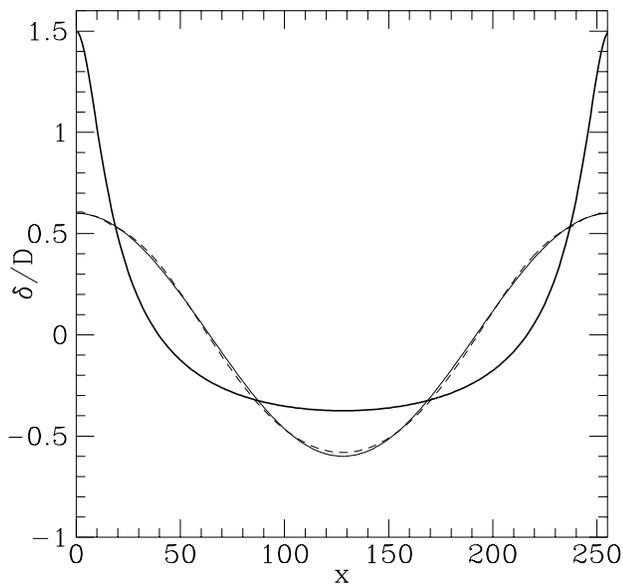

**Figure 1.** Reconstruction of a one-dimensional density field. The current-epoch field (thick curve) is given by (26) with $t_0 = 0.46\tau$ and $\kappa = 2\pi/256$. The order of the truncated series in (16) is $N = 10$. The thin curve shows the exact initial density from (26), while the dashed curve shows the reconstructed initial density. Notice that the evolution predicted by linear theory has been factored out.

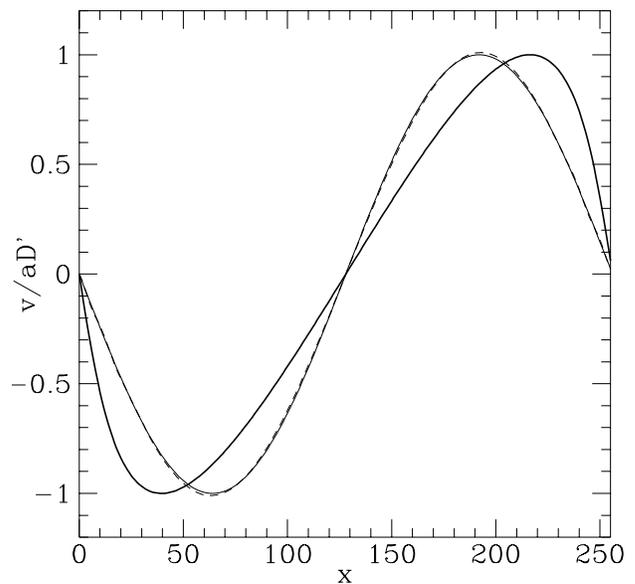

**Figure 2.** The velocity fields associated with the density fields of Fig. 1. The full curves show the initial (thin) and current (thick) fields from (26), while the dashed curves show the corresponding fields obtained from the least-action principle. The current velocity field predicted by the least-action principle is plotted as a dotted curve, but is barely visible because it overlays the analytic curve. The velocity scale is such that the initial velocity amplitude is normalized to unity. The evolution predicted by linear theory has again been factored out.

An accurate description of the sharp peaks and flat-bottomed troughs of $\delta(t_0, x)$ requires significant power in $\delta_{k,0}$ for $k$ large. When the reconstructed initial field is evaluated, this high-$k$ power must be cancelled by an equal amount of high-$k$ power in $\delta_{k,n}$ for $n > 0$, to leave a simple sinusoid. Failure to solve the system (20) accurately, either because one has not iterated to convergence or one has adopted too small a value of $N$, leads to incomplete subtraction of the high-$k$ components of $\delta(t_0, x)$ and small-scale wiggles in the reconstructed fields. The latter are more noticeable in $\delta(0, x)$ than in $v(0, x)$ since the high-$k$ Fourier coefficients of $v$ are smaller than those of $\delta$ by a factor $\sim k^{-1}$.

### 3.2 One-dimensional Gaussian fields

Gaussian random fields are of special interest since they are the simplest random fields, and they arise naturally in the simplest models of the generation of cosmic fluctuations (e.g. Peebles 1993). They are characterized by a power spectrum since the phases of their Fourier modes are uncorrelated with one another. An important geometrical property of a Gaussian field $\delta(x)$ is that it is statistically indistinguishable from $-\delta(x)$. In particular, regions of given over-density have, on the average, the same physical shapes as regions of equivalent under-density. As an initially Gaussian density field evolves under the action of gravity, it ceases to be a Gaussian field since over-dense and under-dense regions evolve differently. In $k$-space this is reflected in the development of correlations between the phases of the field's Fourier components.

Soda et al. (1992) and Gramann (1992) have studied the growth of these correlations in numerical simulations. The interpretation of the results of these calculations is difficult, however, given that phases are determined only modulo $2\pi$.

In this section we investigate the backwards evolution of fields that are Gaussian at the current epoch. In particular, we ask whether a current-epoch Gaussian field can be reached by the action of gravity on a physically realizable initial field. We set

$$\delta_{k,0} \equiv A k^{-2} e^{2\pi i r_k}, \qquad (29)$$

where $0 < r_k \leq 1$ is a random number and the amplitude $A$ is related to the variance in the density, $\sigma^2$, by

$$\sigma^2 = \frac{A^2}{L^2} \sum_k k^{-4}$$
$$= \frac{A^2 L^2}{(2\pi)^4} \zeta(4), \qquad (30)$$

where $\zeta$ is Riemann's zeta function

$$\zeta(m) \equiv \sum_{n=0}^{\infty} n^{-m}; \quad \zeta(4) \approx 1.082. \qquad (31)$$

The full curve in Fig. 3 shows a Gaussian field with Fourier coefficients given by (29), while the dashed curve shows the corresponding initial field from equations (20). The current-epoch field selected shows a void at the centre, and a peak near $x = 0$ (which is identified with $x = 256$). In the initial field, the void is narrower and the peak wider than at the current epoch. *Relative to linear theory*, the void is initially deeper, and the peak lower, than at present. Thus we are here simply seeing the usual tendency of gravity to widen voids and make them flatter-bottomed, whilst raising and sharpening peaks. These trends are apparent in all the realizations of random fields that we have studied.

Fig. 4 shows the velocity fields associated with the density fields of Fig. 3. The velocity changes sign near the extrema of $\delta$, as we expect, given that velocity vectors point away from voids. Similarly, the largest absolute velocities are attained near points at which $\delta$ passes through zero, which shift outward with the expansion of the void.

## 4 TWO-DIMENSIONAL GAUSSIAN FIELDS

The evolution of cosmic fields is much more interesting in two dimensions than in one, since two-dimensional fields display phenomena such as shear and tidal fields that have no one-dimensional analogues (Ellis 1971; Matarrese, Pantano & Saez, et al. 1993; Bertschinger 1993). As Melott & Shandarin (1989) point out, two-dimensional fields have the advantage over three-dimensional ones of being readily displayed and thus providing a useful laboratory in which to study the evolution of cosmic fields.

We have again traced backwards currently Gaussian fields. We take

$$\delta_{\mathbf{k},0} = A k^{-3} e^{2\pi i r_{\mathbf{k}}}, \qquad (32)$$

where the phases $r_{\mathbf{k}}$ are random numbers $0 < r_{\mathbf{k}} \leq 1$. $A$ is related to the variance of the density field by

$$\sigma^2 = \frac{A^2 L}{(2\pi)^6} \sum_{\substack{i,j=0 \\ i=j\neq 0}}^{\infty} \frac{1}{(i^2+j^2)^3} \qquad (33)$$

$$\simeq 4.66 \frac{A^2 L^2}{(2\pi)^6}.$$

The full contours in Fig. 5 are contours of constant density at the current epoch of a field that is now Gaussian with dispersion $\sigma = 1.1$. A peak at center left is balanced by a broad, flat depression that is deepest near $(20, 5)$. The contour $\delta = 0$ is the heavy line. The reconstructed early-time field (dashed for $\delta \geq 0$, dotted for $\delta < 0$) shows the peak to have formed by the amalgamation of two density maxima, one centred on $(10, 17)$ and a smaller one centred on $(25, 16)$. At early times the trough contains two sharp minima at the base of a conical hole. Over time these minima merge and the hole becomes shallower and flatter-bottomed. Thus again we see the tendency of gravity to make troughs broader and shallower, and peaks taller and sharper, as well as its tendency to amalgamate similar features.

Fig. 6 shows three velocity fields associated with the density fields of Fig. 5. The left-hand panel shows the predicted current velocities. As expected, these diverge from the void and converge on the peak. The centre panel shows the reconstructed initial velocity field. This is similar to the current field, the main differences being (i) a shift in the main centre of convergence consequent on a shift in the location of the dominant peak, and (ii) a more complex structure consequent on the existence of a subsidiary peak and trough. The right panel in Fig. 6 shows the velocity field predicted by the Zel'dovich approximation – each arrow in the centre panel has been displaced the appropriate amount, and where this places more than one arrow in a cell, the arrows in that cell have been averaged. Clearly, the velocity field shown in the right panel is a mediocre approximation to the actual current-epoch field shown in the left panel. In particular, distortion by the subsidiary peak is still apparent, and the main point of convergence does not coincide very accurately with the actual density peak. In fact, the middle panel offers at least as good an approximation to the left panel, which suggests that the "frozen-flow" approximation (Matarrese et al. 1992) provides as accurate an estimate of evolved velocity fields as does the more widely employed Zel'dovich approximation.

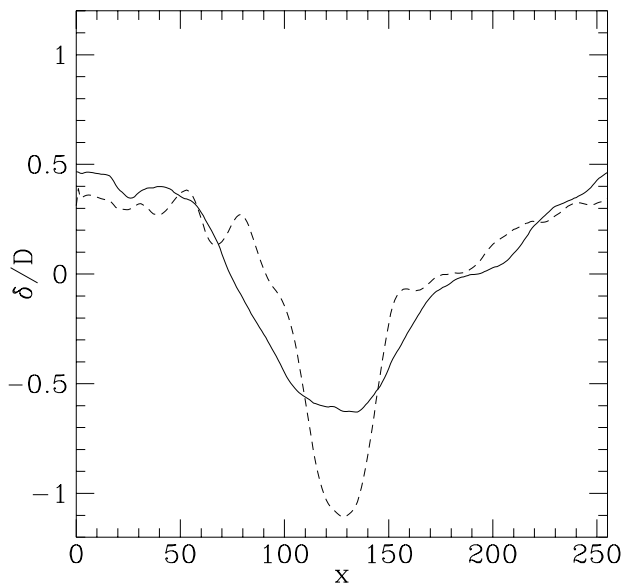

**Figure 3.** Backwards evolution of a current-epoch Gaussian field with Fourier coefficients given by (29) for $\sigma = 0.26$. Full curve: current-epoch density field. Dashed curve: reconstructed initial field. The order of the truncated series (16) is $N = 10$.

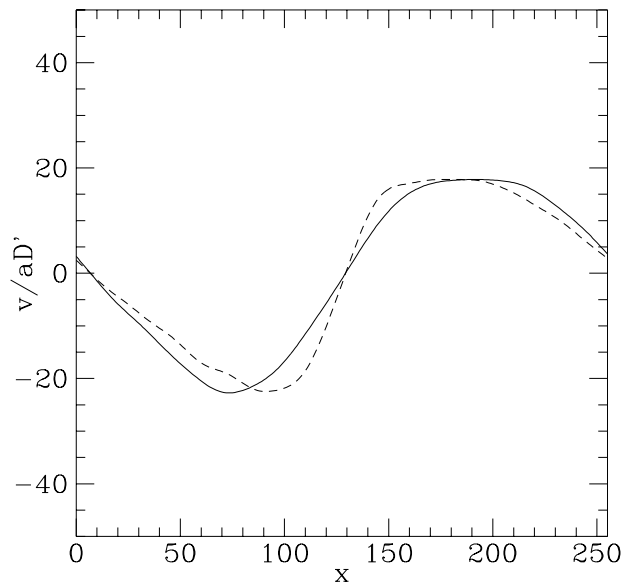

**Figure 4.** Velocity fields corresponding to Fig. 3. Full curve: current-epoch velocity amplitude predicted by equations (20). Dashed curve: reconstructed initial velocity amplitude.

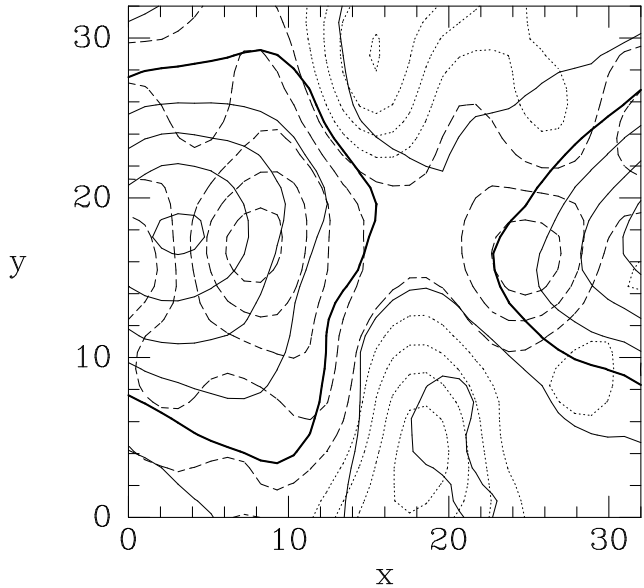

**Figure 5.** Contours of constant present (full) and initial (broken) density for a field which is now Gaussian with Fourier coefficients given by equation (32). The field's dispersion is $\sigma = 1.1$ and the lattice of Fourier modes has $K = 32$ points on a side. Contours are spaced by $\frac{3}{8}$ and contours of negative initial over-density are dotted, while those of zero and positive over-density are dashed. The thick full curve marks the contour on which $\delta = 0$ at the current epoch.

## 5 RELATIONSHIP TO OTHER WORK

In this section we clarify the relation of the present approach to similar work in the literature. Several entirely different expansion schemes for following the evolution of the cosmic fields have been explored. Since these schemes often bear considerable superficial resemblance to one another, it may be felt helpful to summarize their relationship to this work.

The most fundamental distinction is between Eulerian and Lagrangian schemes. Buchert (1992) and Buchert & Ehlers (1993) have explored schemes that are fully Lagrangian in the sense that auxilliary fields are expressed as functions $f(q)$ of the Lagrangian coordinate $q$. Moreover, this work differs from that of Bouchet et al. (1992), Gramann (1993b) and others in that $q$ does not coincide with the Eulerian coordinate $x$ at $t = 0$ but at $t = t_0 > 0$. Consequently, one can write

$$x(t, q) = q + p(t, q), \qquad (34)$$

where $p \to 0$ as $t \to t_0$. The smallness of $p$ is exploited to solve the non-linear equations of motion iteratively, with non-linear terms evaluated using $p$ from the last iteration. The convergence of the series of solutions thus generated is by no means guaranteed. In fact, our own experience with the algebraically similar system (20) suggests that it will converge only for $|t - t_0|$ small. The solutions obtained diverge as $t \to 0$ because they include decaying modes. Consequently, they are incompatible with Proposition II of the Introduction and cannot be used to investigate initial conditions.

A large number of papers employ a similar iterative approach within Eulerian theory (Juszkiewicz (1981); Goroff et al. 1986; Bernardeau 1992; Makino, Sasaki & Suto, 1992; Jain & Bertschinger 1994); the Fourier transforms of the Eulerian equations of motion and continuity are written so that linear terms appear on the left-hand sides and non-linear terms appear on the right. The fields $\delta_k$ etc are then expanded as power series in $D(t)$ and the resulting system of non-linear equations is solved iteratively with the right-hand sides evaluated in advance using the values of the fields from the previous iteration. The first approximation, being obtained by neglecting the right-hand sides, is simply standard linear theory with an arbitrarily chosen amplitude. As in the Lagrangian schemes described above, there is no guarantee that the series obtained converge, and little understanding of the optimum number of terms to employ. The key dif-ferences between this approach and the present work are (i) that we obtain an exact solution of our truncated equations rather than generating a probably asymptotic power series, and (ii) we express the early-time fields as functions of a current-epoch field, rather than expressing the current-epoch fields as functions of the early-time field $\delta(0, x)$. The latter dependence is convenient if one wishes to determine ensemble averages of the fields rather than the evolution of a particular field, and one is willing to assume that the early-time fields are Gaussian. Our approach is the one to choose if one wishes to work with actual, observationally determined, fields.

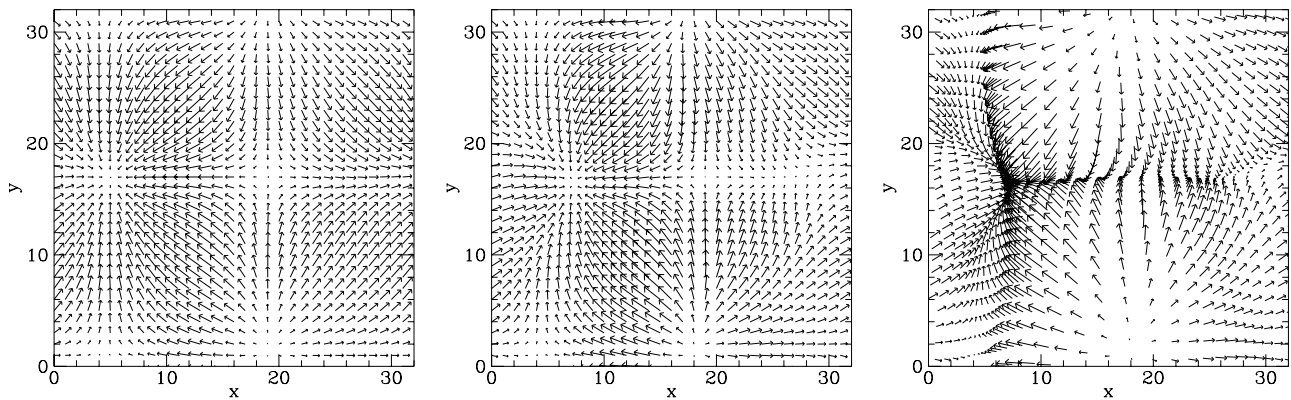

**Figure 6.** Velocity fields corresponding to the density field of Fig. 5. Left: the reconstructed current-epoch velocity field. Centre: the reconstructed initial velocity field. Right: the velocity field predicted by applying the Zel'dovich approximation to the reconstructed initial fields.

As explained in the Introduction, our work differs from that of Bouchet et al. (1992), Gramann (1993a,b) and others in that we seek least-squares approximations to the evolution equations, rather than Taylor-series developments of these solutions.

## 6 CONCLUSIONS

The growth of inhomogeneities from a near-homogeneous early Universe constitutes a two-point boundary-value problem and as such, it is most naturally approached through the least-action principle. The Eulerian formulation of the problem is simpler than the Lagrangian formulation in that: (a) it can be formulated in terms of two unrestricted scalar fields; (b) in it the gravitational potential is simply related to one of these fundamental fields; (c) its predictions for the current epoch may be easily and directly compared with observations. The principal weakness of the Eulerian approach is that the density field becomes singular as soon as caustics cross, with the result that the radius of convergence of any Taylor series for the density contrast is probably small. Fortunately, when the variational constraint equations are employed, the power series involved are least-squares fitted to the solutions rather than established as Taylor series, with the result that their validity is not as narrowly restricted as when one approaches the problem from the differential equations of motion.

In Section 2 we have formulated the application of the least-action principle to cosmic inhomogeneities in terms of coefficients $\delta_{\mathbf{k},n}$ and $\alpha_{\mathbf{k},n}$ in expansions of the Fourier amplitudes $\delta_{\mathbf{k}}(t)$ and $\alpha_{\mathbf{k}}(t)$ in powers of $(D-1)$ ($n = 0, 1, \ldots$), where $D$ is the linear growth factor and $\alpha(t, x)$ is the velocity potential. These coefficients solve a non-linear system of equations (20). In these equations one may consider either the amplitudes $\delta_{\mathbf{k},0}$ or the amplitudes $\alpha_{\mathbf{k},0}$ to play the rôle of boundary conditions, which specify the structure of the cosmic fields at the present time. The solutions to the equations then yield the structure of the conjugate field ($\alpha$ or $\delta$, respectively) at the current epoch as well as the structure of both fields at all earlier epochs.

In Section 3 we have solved the system (20) for the one-dimensional case, for which an exact solution is known. We obtain solutions by increasing the magnitude of the non-linear terms in the equations to their true values in stages, and iterating a trial solution to a fully converged solution at each stage. This technique enables us to solve equations (20) for any meaningful values of the current-epoch fields.

The calculated backwards evolution of the exact solution agrees well with the form of the solution at earlier times provided sufficient ($N \gtrsim 3$) powers of $(D-1)$ are employed. A field whose Fourier representation involves significant power at large wavenumbers $k$ evolves backwards into a pure sinusoid in consequence of the high-$k$ components of $\delta(t_0, x)$ being more and more exactly cancelled by the high-$k$ coefficients of $(D-1)^n$ for large $n$. Errors in the completeness of this cancellation determine the exactitude with which early-time fields can be determined. The early-time and current-epoch velocity fields can be more accurately determined because they involve less high-$k$ power.

We have used our machinery to trace backwards in one and two dimensions fields that are currently Gaussian. We find that at early times their peaks are lower and flatter topped, and their troughs deeper and more pointed. In two dimensions this exercise reveals the tendency of structures to form by the merging of sub-structures. Hence, on tracing a peak backwards in time in Section 4, we found it to have formed by the merging of two smaller peaks that were initially located on either side of it, while a broad shallow trough had emerged from the amalgamation of two rather vee-shaped holes.

Since our reconstruction scheme yields both the current- and the early-epoch velocity fields, it enables us to test the validity of the Zel'dovich approximation. In Section 4 we concluded that the Zel'dovich approximation does not provide a better approximation than the "frozen-flow" approximation of Matarrese et al. (1992).

Section 5 clarified the relation of this work to similar expansion schemes that have been employed to evolve the cosmic fields forwards in time. The essential differences are: (i) while we approximate solutions of the evolution equations by polynomials in $(D-1)$, other schemes generate power-series approximations that are either Taylor series or formal power series that are probably at best asymptotic; (ii) we express the early-time fields as functions of the current-epoch fields, while in many other schemes the functional dependence is reversed.

Before the reconstruction scheme developed here can be applied to real data, two things have to be done: (i) the scheme has to be implemented in $d = 3$, (ii) it has to be interfaced with an effective means of determining either $\delta_{\mathbf{k},0}$ or $\alpha_{\mathbf{k},0}$. Implementation in $d = 3$ is straightforward; the only drawbacks to working in three dimensions are the difficulty of displaying the results and the greatly increased computational cost – both the memory and the CPU time required are proportional to $K^d$. In fact, we have already tested a three-dimensional implementation of the present scheme.

The observational determination of either $\delta_{\mathbf{k},0}$ or $\alpha_{\mathbf{k},0}$ is harder. The $\alpha_{\mathbf{k},0}$ are the easiest quantities to estimate, since every distance to a galaxy of known redshift yields an estimate of $\hat{r}_g \cdot \nabla \alpha$, where $\hat{r}_g$ is the direction to that galaxy. An attractive scheme is to choose the $\alpha_{\mathbf{k},0}$ which minimize

$$\chi^2 = \sum_{\text{galaxies } g} \left[ v_g - \mathrm{i} \frac{\dot{D}(t_0)}{L^3} \hat{r}_g \cdot \sum_{\mathbf{k}} \mathbf{k}\, \alpha_{\mathbf{k},0}\, \mathrm{e}^{\mathrm{i} \mathbf{k} \cdot \mathbf{x}_g} \right]^2, \quad (35)$$

where $x_g$ and $v_g$ are the position and radial peculiar velocity of the $g$th galaxy in some sample. Since this is a standard linear least-squares problem, its solution would be straightforward if one's galaxy sample contained more galaxies than a reasonable grid in $\mathbf{k}$-space does points ($\gtrsim 30,000$). In practice, one is unlikely to be able to observationally determine so many coefficients $\alpha_{k,0}$ in the near future, and it will be necessary to set $\alpha_{k,0} = 0$ for $k > k_{\max}$, say. Nonetheless, for $n > 0$ one's calculations should include coefficients $\alpha_{k,n}$ for $k > k_{\max}$, since as one traces the fields backwards in time, power will appear at these high frequencies even if none is present in $\alpha(x)$ at the current epoch. We hope to report on an investigation along these lines in the not too distant future.

## ACKNOWLEDGMENTS

The authors thank George Efstathiou, Cedric Lacey and Will Saunders for useful discussions. MPS acknowledges receipt of a studentship from the Basque government's *Hezkuntza, Ikerketa eta Unibertsitate Saila*.

## APPENDIX A: EVALUATION OF ANGLE BRACKETS

We evaluate for the special case $\Omega = 1$ the integrals over time required in equations (13) and (14). By the definition (15) of $\langle x \rangle$, they are

$$\langle \dot{f}_n g_r \rangle = \int_0^{t_0} dt\, a(t)^2\, \dot{f}_n(t)\, g_r(t),$$

$$\langle g_n g_r \rangle = \int_0^{t_0} dt\, a(t)^2\, g_n(t)\, g_r(t),$$

$$\langle f_n g_r g_m \rangle = \int_0^{t_0} dt\, a(t)^2\, f_n(t)\, g_r(t)\, g_m(t), \quad (A1)$$

$$\langle \rho_b f_n f_r \rangle = \int_0^{t_0} dt\, a(t)^2\, \rho_b(t)\, f_n(t)\, f_r(t),$$

$$\langle (\dot{g}_n + 2\frac{\dot{a}}{a} g_n) f_r \rangle = \int_0^{t_0} dt\, a(t)^2 \left[ \dot{g}_n(t) + 2\frac{\dot{a}}{a} g_n(t) \right] f_r(t),$$

where

$$a(t) = \left( \frac{t}{t_0} \right)^{2/3} \quad \text{and} \quad \rho_b(t) = \frac{1}{6\pi G t^2}. \quad (A2)$$

We use the result $D = a = (t/t_0)^{2/3}$ to effect the replacement

$$\int_0^1 dt\, a^2 \quad \to \quad \tfrac{3}{2} t_0 \int_0^1 dD\, D^{5/2}. \quad (A3)$$

Next use the definitions (11) to evaluate $f_n$, $g_n$ and their time derivatives in terms of $D$. This yields

$$\langle \dot{f}_n g_r \rangle = \frac{2}{3 t_0} \left[ I(\tfrac{3}{2}, n+r) + n I(\tfrac{5}{2}, n+r-1) \right],$$

$$\langle g_n g_r \rangle = \frac{2}{3 t_0} I(\tfrac{3}{2}, n+r),$$

$$\langle f_n g_m g_r \rangle = \frac{2}{3 t_0} I(\tfrac{5}{2}, m+n+r), \quad (A4)$$

$$\langle \rho_b f_n f_r \rangle = \frac{1}{4\pi G t_0} I(\tfrac{3}{2}, n+r),$$

$$\langle (\dot{g}_n + 2\tfrac{\dot{a}}{a} g_n) f_r \rangle = \frac{1}{t_0} \left[ I(\tfrac{3}{2}, n+r) + \tfrac{2}{3} n I(\tfrac{5}{2}, n+r-1) \right],$$

where for $n = 0, 1, 2, \ldots$ and $\alpha \geq 0$

$$I(\alpha, n) \equiv \int_0^1 du\, u^\alpha\, (u-1)^n$$
$$= (-1)^n\, n! \prod_{i=1}^{n+1} (i+\alpha)^{-1}. \quad (A5)$$

(The proof of the last equality is by induction on $n$ from $n = 0$ by repeated integration by parts.) It follows that

$$I(\tfrac{5}{2}, n-1) = -\frac{5}{2n} I(\tfrac{3}{2}, n). \quad (A6)$$

Since

$$\prod_{i=1}^{n+1} (i + \tfrac{3}{2})^{-1} = 4!\, 2^{2n}\, \frac{(n+2)!}{(2n+5)!}, \quad (A7)$$

we have finally

$$I(\tfrac{3}{2}, n) = 4!\phi(n) , \qquad I(\tfrac{5}{2}, n - 1) = \frac{-5!}{2n}\phi(n), \tag{A8}$$

where

$$\phi(n) \equiv (-4)^n \frac{n!(n+2)!}{(2n+5)!}. \tag{A9}$$

## APPENDIX B: CONSTRAINT EQUATIONS

On substituting from Appendix A for the values of the angle brackets, truncating the sums over $n$, and replacing the Fourier integrals by Fourier sums as described in the text, equations (13) and (14) become

$$\sum_{n=0}^{N} \phi(n+r)\left[\left(1 - \tfrac{5}{2}\frac{n}{n+r}\right)\delta_{\mathbf{k},n} - k^2 \alpha_{\mathbf{k},n}\right]$$
$$= \sum_{n,m} -\tfrac{5}{2}\frac{\phi(m+n+r+1)}{m+n+r+1} \times$$
$$\times \frac{1}{L^d}\sum_{\mathbf{p}} \mathbf{k}\cdot\mathbf{p}\, \delta_{\mathbf{k}-\mathbf{p},n}\, \alpha_{\mathbf{p},m},$$
$$\sum_{n=0}^{N} \phi(n+r)\left[\delta_{\mathbf{k},n} - k^2\left(1 - \frac{5n}{3(n+r)}\right)\alpha_{\mathbf{k},n}\right] \tag{B1}$$
$$= \sum_{m,n} \frac{5k^2}{6}\frac{\phi(m+n+r+1)}{m+n+r+1} \times$$
$$\times \frac{1}{L^d}\sum_{\mathbf{p}}(\mathbf{k}-\mathbf{p})\cdot\mathbf{p}\, \alpha_{\mathbf{k}-\mathbf{p},m}\, \alpha_{\mathbf{p},n}.$$

The sum over $\mathbf{p}$ in the first of equations (B1) can be rewritten as a convolution since

$$\sum_{\mathbf{p}} \mathbf{k}\cdot\mathbf{p}\, \delta_{\mathbf{k}-\mathbf{p}}\, \alpha_{\mathbf{p}} = \sum_{\mathbf{p}}\left[(\mathbf{k}-\mathbf{p})\cdot\mathbf{p}\, \delta_{\mathbf{k}-\mathbf{p}}\, \alpha_{\mathbf{p}} + \delta_{\mathbf{k}-\mathbf{p}}\alpha_{\mathbf{p}}p^2\right]. \tag{B2}$$

When we confine the Fourier sums to a sublattice of $K^d$ points, these convolutions can be evaluated efficiently, and without further approximation, by application of the discrete Fourier transform theorem.

This paper has been produced using the Blackwell Scientific Publications T<sub>E</sub>X macros.